\documentclass[journal, 10pt]{IEEEtran}
\usepackage[utf8]{inputenc}

\usepackage{amsmath,amssymb,amsfonts,amsthm}
\usepackage{algorithmic}
\usepackage{algorithm}
\usepackage{graphicx}
\usepackage{xcolor}
\usepackage{mathtools}
\usepackage{multirow}
\usepackage{caption}
\usepackage[font = footnotesize]{subfig}
\usepackage{tikzpagenodes}

\theoremstyle{remark}

\usepackage[style=ieee, backend=biber]{biblatex}
\title{Semantic Communication with Conceptual Spaces}

\author{
    Dylan Wheeler, \IEEEmembership{Student Member, IEEE,} Erin E. Tripp, and Balasubramaniam Natarajan, \IEEEmembership{Senior Member, IEEE}
    \thanks{This work was funded by Air Force Office of Scientific Research grant 21RICO035. Any opinions, findings and conclusions or recommendations expressed in this material are those of the authors and do not necessarily reflect the views of the U.S.\ Air Force Research
Laboratory. Cleared for public release 29 August 2022: case number AFRL-2022-4130.}
    \thanks{D. Wheeler and B. Natarajan are with the Mike Wiegers Department of Electrical and Computer Engineering at Kansas State University (email: dylan84@ksu.edu)}
    \thanks{ E. E. Tripp is with the Air Force Research Laboratory Information Directorate (email: erin.tripp.4@us.af.mil)}
}

\begin{document}

\maketitle

\begin{tikzpicture}[remember picture,overlay]
    \node[align=center] at ([yshift=2em]current page text area.north) {\footnotesize This article has been accepted for publication in IEEE Communications Letters. This is the author's version which has not\\ \footnotesize been fully edited and content may change prior to final publication. Citation information: DOI 10.1109/LCOMM.2022.3230246};
\end{tikzpicture}%

\begin{tikzpicture}[remember picture,overlay]
    \node[align=center] at ([yshift=-3em]current page text area.south) {\scriptsize \textcopyright 2022 IEEE.  Personal use of this material is permitted.  Permission from IEEE must be obtained for all other uses, in any current or future media, including reprinting/republishing this \\ \scriptsize material for advertising or promotional purposes, creating new collective works, for resale or redistribution to servers or lists, or reuse of any copyrighted component of this work in other works.};
\end{tikzpicture}%

\vspace{-.72cm}

\begin{abstract}
Despite the fact that Shannon and Weaver's Mathematical Theory of Communication was published over 70 years ago, all communication systems continue to operate at the first of three levels defined in this theory: the technical level. In this letter, we argue that a transition to the semantic level embodies a natural, important step in the evolution of communication technologies. Furthermore, we propose a novel approach to engineering semantic communication using conceptual spaces and functional compression. We introduce a model of semantic communication utilizing this approach, and simulate communication of image semantics demonstrating a 99.79\% reduction in rate.
\end{abstract}


\vspace{-0.4cm}

\section{Introduction}
\label{sec_intro}

While the world is only a few years into the deployment of fifth generation (5G) communication networks, researchers have set their sights on what might come next \cite{jiang_road6G_2021}. Often referred to as the beyond-5G (B5G) or sixth generation (6G) network, some visions have emerged of technologies that have the potential to play a large role in this network, including pervasive artificial intelligence (AI), and semantic communication \cite{strinati_6Gfrontier_2019}. Confronted with the challenge of an exponential rise in global data traffic \cite{ericsson_traffic_2021}, we believe that semantic communication provides a promising approach to address this challenge.

In Shannon and Weaver's groundbreaking work \cite{shannon_weaver}, they define three fundamental communication problems:
\begin{enumerate}
    \item[A.] \textit{Technical}: how to accurately transmit symbols?
    \item[B.] \textit{Semantic}: how to accurately convey meaning?
    \item[C.] \textit{Effective}: how to affect conduct in the desired way?
\end{enumerate}
The focus of their work is on the technical problem. Indeed, it is stated that ``these semantic aspects of communication are irrelevant to the engineering problem.'' Consequently, all communication systems today operate at this first level (we also refer to this as the syntactic level). While adequate for faithfully conveying information, this focus brings with it inefficiency stemming from the fact that a syntactic error does not necessarily induce a semantic error. For example, suppose a sender transmits the sentence ``All communication is semantic,'' but the receiver is presented with ``All comnunication is semantic.'' Erroneous transmissions are often addressed by retransmission schemes or error-correcting codes, requiring additional valuable resources. However, viewed from the semantic level, this is a minor error and most likely will preserve the semantic information intended by the transmitter. 

From this simple example, we observe that semantic-oriented communication may be more robust to errors and, in turn, more efficient than traditional systems. Then the question is this: \textit{how can we engineer semantics in a communication system?} Looming over this issue is an even more fundamental question: \textit{how should we define semantics?} For the latter, we adopt the \textit{conceptual space} theory of semantics proposed by Peter G\"ardenfors \cite{gardenfors_conceptSpace_2000, gardenfors_semanticTheory_2014}. This geometric representation of meaning can provide the foundation for efficient semantic communication systems. To address the former question, we propose a system with a semantic encoder/decoder based on this theory. Moreover, we propose the use of \textit{functional compression} to optimize the semantic system, in which data from one or more sources is compressed such that a function of this data can be accurately computed \cite{doshi_functional_2010}. We show that encoding data to preserve semantic information can be viewed as a particular example of functional compression.

\subsection{Related Work}
\label{subsec_rel_work}

There has been a recent surge of interest in semantic communication, sparked by the success of modern AI. One of the most prominent approaches is termed \textit{Deep-SC}, which uses deep learning (DL) to ``learn'' the semantics of text communication \cite{xie_deepsc_2021}. Another idea regards semantics as the \textit{significance} of information \cite{uysal_semcomm_2021}, where the meaning of information is captured by metrics like age of information and value of information. Lan \textit{et al.} provide a review of many of the recent works in the field of semantic communication \cite{lan_21}.

Functional compression is a generalization of the well-known source coding problem \cite{shannon_weaver}. In \cite{doshi_functional_2010}, Doshi et al. extend this work to general functions of two discrete, finite sources. The problem is represented as a graph coloring problem, with feasibility conditions and optimality bounds derived. While this can be solved exactly for some nice functions, this problem is NP-hard in general. 




Since G\"ardenfors' proposal of conceptual spaces in 2000, there have been some works that attempt to extend fundamentals of the proposed theory. Rickard, Aisbett, and Gibbon propose the addition of fuzzy logic to the theory and formalize many of the abstract concepts in \cite{rickard_knowledge_2007}, and extend this work to type-2 fuzzy spaces in \cite{rickard_type-2_2010}. In addition, some have used conceptual spaces to facilitate a diverse set of applications, from robot learning \cite{cubek_high-level_2015} to space event characterization \cite{chapman_conceptual_2020}.

\subsection{Contributions}
\label{subsec_contri}

In this letter, for the first time, we show how the theory of conceptual spaces can be deployed to develop a semantic communication system and demonstrate the benefits of such a system. The main contributions of this letter include:
\begin{itemize}
    \item A novel approach to designing semantic communication systems based on conceptual space semantics
    \item The introduction of a formal approach to the study of such systems using functional compression techniques
    \item Confirmation of the intuitive benefits of semantic communication through simulation of image transmission, demonstrating faithful transmission of meaning with 99.79\% reduction in rate
\end{itemize}


\section{The Geometry of Meaning}
\label{sec_theory}

In this section, we attempt to answer the latter of the two questions posed in Section \ref{sec_intro}: \textit{how should we define semantics?} 
Proposed as a cognitive model of the way humans conceptualize ideas, conceptual spaces offer a geometric view of concepts.

Summarizing the key components introduced in \cite{gardenfors_conceptSpace_2000} and \cite{gardenfors_semanticTheory_2014}, a \textit{conceptual space} is essentially the knowledge base of an agent. A conceptual space can be characterized as a collection of \textit{domains}; each domain is then made up of integral \textit{quality dimensions} that define the geometry of the domain. Based on these definitions, a \textit{property} is defined as a convex region of a single domain, and a \textit{concept} is defined as a collection of regions across domains within the conceptual space.

\begin{figure}[b]
    \centering
    \includegraphics[scale = 0.27]{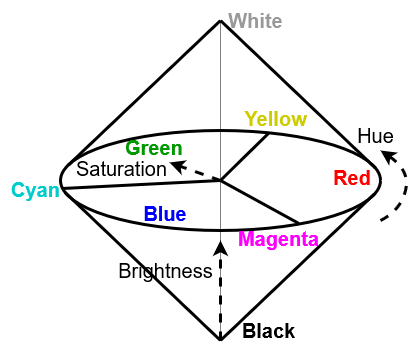}
    \caption{Color domain. Dashed lines indicate the three dimensions}
    \label{fig_spindle}
\end{figure}

To illustrate, consider the example of the color domain. Studies indicate that humans perceive three dimensions regarding color: hue, saturation, and brightness \cite{gardenfors_conceptSpace_2000}. 
These dimensions form the color spindle shown in Figure \ref{fig_spindle}. An example property could be \textit{red}, which corresponds to the convex region of the spindle with a red hue value. Another property is \textit{dark}, which describes colors in the bottom half of the spindle. Now add a second domain to the space, e.g. the shape domain. In this domain, we have a convex region corresponding to \textit{cube}. Then the \textit{red} region of the color domain with the \textit{cube} region of the shape domain define the concept of a \textit{red cube}.

The major implication of this theory for semantic communication is the formalization of \textit{meaning as a geometric concept}. With this formalization, semantic similarity becomes a simple distance measure (provided a metric exists on the domains), and semantic error can be easily quantified. Indeed, this framework elegantly captures the idea that a syntactic error does not always induce a semantic error. When communicating an idea (a point in the conceptual space), perhaps a syntactic error moves the idea to a different point in the space; as long as the point remains within the regions defined by the concept, a semantic error will not occur. 

\section{Making Communication Semantic}
\label{sec_methods}

With our definition of semantics in place, we now turn to the first question posed in Section \ref{sec_intro}: \textit{how can we engineer semantic communication?}

\subsection{The General Model}
\label{subsec_gen_model}

\begin{figure}[t]
    \centering
    \includegraphics[scale = 0.492]{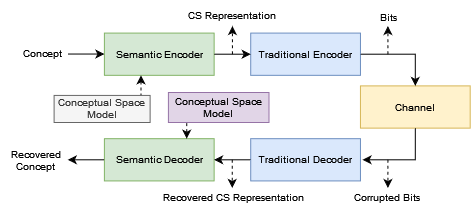}
    \caption{General Model of Semantic Communication with Conceptual Spaces}
    \label{fig_semcomm_model}
\end{figure}

The block diagram of our proposed approach is given in Figure \ref{fig_semcomm_model}. Formally, suppose there is some concept to be communicated, denoted by $\textbf{z}$. The system begins with some data $\textbf{x}$ describing $\textbf{z}$, which is first passed through the \textit{semantic encoder}; this is represented by some functional mapping $\textbf{p} = e(\textbf{x})$. The function $e$ can be thought as an \textit{embedding} function, which maps the description to some point $\textbf{p}$ in the conceptual space, where $\textbf{p}$ is a vector of coordinates. This mapping can by either analytically derived or learned with a neural network. The coordinates $\textbf{p}$ are then used as the input to the traditional encoder, which performs functions such as quantization, source coding, etc. and outputs a string of bits $\textbf{b}$; in functional terms, we have $\textbf{b} = g(\textbf{p})$. 
The semantic source can then be thought of as a composition of functions,
\begin{equation}
    \textbf{b} = g(e(\textbf{x})).
\end{equation}

The receiver begins with an input of bits $\hat{\textbf{b}}$ which are potentially distorted by the communication channel, i.e., $\hat{\textbf{b}} = h(\textbf{b})$. These bits serve as the input to the traditional decoder, which performs standard functions such as source and channel decoding. Denote these functions by $f$, so that we have $\hat{\textbf{p}} = f(\hat{\textbf{b}})$, where $\hat{\textbf{p}}$ represents the (possibly distorted) coordinates of the conceptual space. Finally, let $d$ denote the function de-mapping the conceptual space representation to the concept it represents. Then the semantic receiver can be written as
\begin{equation}
    \hat{\textbf{z}} = d(f(\hat{\textbf{b}})),
\end{equation}
and the end-to-end semantic system is written as
\begin{equation}
    \hat{\textbf{z}} = d(f(h(g(e(\textbf{x}))))) = \psi(\textbf{x}).
    \label{eq_end_to_end}
\end{equation}

If $\hat{\textbf{b}} \neq \textbf{b}$, we say a \textit{syntactic error} has occurred. If $\hat{\textbf{z}} \neq \textbf{z}$, then we call this a \textit{semantic error}. Note that $\hat{\textbf{b}} = \textbf{b}$ (perfect syntactic communication) yields the minimum probability of semantic error. Conversely $\hat{\textbf{b}} \neq \textbf{b}$ does not necessarily imply a semantic error, matching our intuition of a semantic communication system. Therefore, with the goal of accurate semantic communication, \textit{it may be possible to relax the requirements of the technical system while meeting the requirements of the semantic system.} First observe that $g,h$, and $f$ account for the technical accuracy, which in turn will impact the semantic accuracy to some degree. Moreover, the semantic accuracy also depends on the ability of $e$ to extract semantic representations from the initial data, and on the ability of $d$ to recover concepts from potentially distorted semantic representations. The benefit of using conceptual space representations is in the ability to simply quantify this semantic distortion as a distance in the conceptual space; let this distance be denoted by $\delta(\textbf{p}_i, \textbf{p}_j)$. If we consider $\textbf{p}^*$ as the central point, or prototype, of the true concept (recall a concept is composed of convex regions), then we can characterize the semantic distortion incurred throughout the overall system as
\begin{equation}
    \delta(\textbf{p}^*, \hat{\textbf{p}}) = \Delta(\delta(\textbf{p}^*, \textbf{p}), \delta(\textbf{p}, \hat{\textbf{p}})),
\end{equation}
where $\Delta$ is a general function of the distortions resulting from semantic encoder and syntactic error. If we further assume $\delta$ to be a metric satisfying the triangle inequality, we can write
\begin{equation}
    \delta(\textbf{p}^*, \hat{\textbf{p}}) \leq \delta(\textbf{p}^*, \textbf{p}) + \delta(\textbf{p}, \hat{\textbf{p}}).
    \label{eq_sem_dist}
\end{equation}
Thus, by bounding the two right-hand terms in (\ref{eq_sem_dist}), we can effectively bound the overall semantic distortion and ensure reliable semantic communication. Moreover, if the semantic encoder is known to have low distortion, we can relax constraints on the technical system while maintaining an acceptable level of overall semantic distortion, and vice versa.

We recognize the fact that \textit{general} semantic communication may require a complex conceptual space, and obtaining such a model is a research challenge. We believe that semantic communication with conceptual spaces will have the largest initial impact in task-specific, goal-oriented scenarios, where the domain knowledge is relatively simple to model.


\subsection{Toward Functional Compression}
\label{subsec_optimizing}




As with traditional communication, we would like to optimize semantic communication to be as efficient as possible through data compression. The objective of traditional source coding is to compress data for exact reconstruction:
\begin{equation*}
    \min_{g,f} E[\ell] \quad \text{ s.t. } \quad \ell = \text{length}(g(\textbf{x})) \: \text{ and } \: f(g(\textbf{x})) = \textbf{x},
\end{equation*}
where length$(g(\textbf{x}))$ counts the number of (typically binary) symbols used to encode $\textbf{x}$, and $E[\cdot]$ denotes expectation. Functional compression seeks to compress data with respect to a function:
\begin{equation*}
    \min_{g,f} E[\ell] \quad \text{ s.t. } \quad \ell = \text{length}(g(\textbf{x})) \: \text{ and } \: f(g(\textbf{x})) = \varphi(\textbf{x}).
\end{equation*}

In particular, this can be done by identifying equivalence classes of the target function $\varphi$ to eliminate redundancies in the encoded data. An equivalence class is a subset of the input space for which $\varphi$ remains constant; or, from the point of view of the receiver, a set of inputs which are indistinguishable. For example, suppose $\varphi(x) = x \mod 2$ for $x$ drawn uniformly from $\{0, 1, 2, 3\}.$ Traditionally, 2 bits are needed to represent each value, but by identifying the equivalence classes $\{0, 2\}$ and $\{1, 3\}$, this can be reduced to 1 bit.

As mentioned above, this problem is NP-hard in general. Here, we propose approaching the problem by first taking a step back. Rather than considering general functions, we consider high-level tasks and then model these tasks using conceptual spaces. 
For the semantic communication system, we can define the following functional compression problem:
\begin{equation}
    \min_{e,g,f,d} E[\ell] \quad \text{ s.t. } \quad \ell = \text{length}(\textbf{b}) \: \text{ and } \: \psi(\textbf{x}) = \textbf{z}.
    \label{eq_fc_semantic_concept}
\end{equation} 
In particular, by assuming a minimum-distance decoder $d$ on the underlying conceptual space, (\ref{eq_fc_semantic_concept}) can be equivalently written as
\begin{equation}
    \min_{e,g,f} E[\ell] \quad \text{ s.t. } \quad \ell = \text{length}(\textbf{b}) \: \text{ and } \: \delta(\textbf{p}^*, \hat{\textbf{p}}) \leq \tau,
    \label{eq_fc_semantic_distance}
\end{equation} 
where $\tau$ is some distortion threshold.
Constructing the target function with conceptual spaces allows us to implicitly include equivalence relations based on the concepts within the space, for which efficient semantic encodings can be obtained. We leave further exploration of this problem to future work.

\vspace{-.1cm}

\section{Experimental Results}
\label{sec_results}


We examine semantic communication of images using the German Traffic Sign Recognition Benchmark dataset \cite{xiang_gtsrb_dataset}, inspired by an autonomous driving application. For our tests, we define a conceptual space with two domains, namely the color domain (Figure \ref{fig_spindle}) and the domain of regular polygons. For the polygon domain, we consider a single dimension quantified by the ratio $r$ of the maximum distance from the center of a shape to its boundary to the minimum of such distance. For a triangle $r=2$ and for a circle $r=1$. Concepts considered in this experiment are: yellow square, red triangle, red octagon, red circle, and blue circle. For example, the prototype ratio, hue, saturation, and brightness values for ``yellow square'' are
\vspace{-.09cm}
\begin{align*}
    \textbf{p}^*_{\text{y.s.}} 
    &= [1.4142 \:\:\: 0.1667 \:\:\: 1 \:\:\: 0.9714]'.
\end{align*}
\vspace{-.5cm}

We wish to compare the proposed system to a more traditional system. For the semantic system, we train a convolutional neural network (CNN) with three convolutional layers and two fully connected layers for a total of 1,692 parameters to learn the semantic encoder function $e$ from the image space to the conceptual space. As such, the prototype values of each concept are used as the training labels. Since hue is a circular dimension, a custom loss function is required to capture distance within the space. Denoting the coordinates of the space as $\textbf{p} = (r \:\: h \:\: s \:\:b)'$, the loss function of the CNN is defined as
\begin{equation}
    \delta(\textbf{p},\hat{\textbf{p}}) = \frac{1}{4}\left((r - \hat{r})^2 + (s - \hat{s})^2 + (b - \hat{b})^2 + \gamma(h,\hat{h})^2\right),
\end{equation}
where
\begin{equation*}
    \gamma(h,\hat{h}) = -\frac{1}{\rho} \ln\left( \frac{1}{2}e^{-\rho \vert h - \hat{h}\vert} + \frac{1}{2} e^{-\rho(1-\vert h-\hat{h} \vert)} \right)
\end{equation*}
is an approximation of the circular distance given by $\min(\vert h-\hat{h} \vert, \vert 1 - (h-\hat{h})\vert)$ with parameter $\rho > 0$. The output of the semantic encoder is then quantized to $n_b$ bits/value to form a $4n_b$-bit packet, which is modulated using binary phase-shift keying (BPSK) and transmitted over a Rayleigh fading channel. At the receiver, BPSK demodulation is performed and a minimum-distance decoder is used to recover the concept.

We also simulate a traditional system which transmits the entire $25\times25$-pixel image over the channel. Each pixel is represented as three $n_b$-bit values, which undergo identical modulation and channel effects as above. At the receiver, a CNN with identical structure to that of the semantic system classifies the concept from the recovered image. This network is trained using the classic cross-entropy loss function, using one-hot encoded vectors as labels.

\begin{figure}[htbp]
    \centering
    \subfloat[\label{fig_semVsSynt}]{%
       \includegraphics[scale = 0.54]{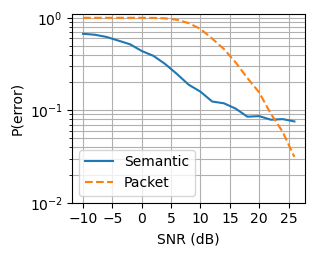}
    }
    \hfill
    \subfloat[\label{fig_semVsBasic}]{%
        \includegraphics[scale = 0.54]{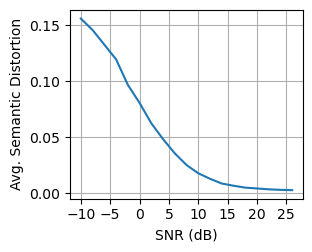}
    }
    \caption{Simulation results for the conceptual space-based semantic communication system. (a) comparison of semantic and syntactic probabilities of error and (b) average end-to-end semantic distortion $\delta(\textbf{p}^*, \hat{\textbf{p}})$}
    \label{fig_results}
\end{figure}

First, we demonstrate that a syntactic error does not necessarily induce a semantic error. We define a semantic error as an incorrectly recovered concept ($\hat{\textbf{z}} \neq \textbf{z}$), whereas a syntactic error occurs when the transmitted packet contains one or more bit errors ($\hat{\textbf{b}} \neq \textbf{b}$). Figure \ref{fig_results}(a) shows the performance for the semantic system with $n_b = 8$ when the signal-to-noise ratio (SNR) is varied. It is observed that the system is able to perform well \textit{semantically} despite significant syntactic errors; for example, at 15dB, approximately 40\% of transmitted packets contain errors, but the meaning is still accurately conveyed 90\% of the time. As the channel conditions improve, the performance becomes limited at an SNR of approximately 20dB due to the nonzero semantic distortion floor illustrated in Figure \ref{fig_results}(b). By improving the semantic encoder, the high-SNR performance of the overall system will improve accordingly.


\vspace{-.5cm}

\vspace{-.2cm}

Next we compare the performance of the two systems as a function of the communication rate; these results are given in Table \ref{tab_results}. Here, the rate of the semantic system is 4$n_b$ bits/transmission, while the rate of the traditional system is $(25\times25\times3)n_b$ bits/transmission, and $n_b$ is varied to obtain different rate values. Thus, the semantic system achieves a rate reduction of $99.79\%$ over the traditional system. Furthermore, we observe that the semantic system is able to make better use of fewer bits, as the meaning conveyed is significantly more accurate at lower values of $n_b$ than for the traditional system. Conversely, as the number of quantization bits increases, the traditional system is shown to outperform the semantic system at the cost of dramatically greater communication rate. 


\section{Conclusion}
\label{sec_conclude}

In this letter, we introduce a novel approach to semantic communication through the use of conceptual space theory. 
We describe the model of a semantic communication system under this theory, and propose functional compression as a method of obtaining optimal encoding schemes for the semantic system. Through simulation of image transmission, we provide quantitative results illustrating the ability of the semantic communication system to faithfully convey meaning with a massive reduction in communication rate.

In the future, we intend to utilize the mathematical foundations of functional compression to provide optimality results for the proposed system. In addition, we plan to build off of the methods introduced in \cite{rickard_knowledge_2007} to incorporate fuzzy techniques into semantic communication.

\begin{table}[t]
    \centering
    \begin{tabular}{|c|c|c|c|c|c|}\cline{2-5}
     \multicolumn{1}{c}{} \vline  & \multicolumn{2}{c}{\textit{Semantic System}} \vline & \multicolumn{2}{c}{\textit{Traditional System}} \vline & \multicolumn{1}{c}{} \\\hline
     $n_b$  & \textit{Rate}    & $P(\hat{\textbf{z}} \neq \textbf{z})$     & \textit{Rate}   & $P(\hat{\textbf{z}} \neq \textbf{z})$ & \textit{Rate Reduction}    \\\hline
     2   & 8    & 17.1\%    & 3750    & 70.9\%  & \multirow{3}{*}{99.79\%} \\\cline{1-5}
     5   & 20   & 10.7\%    & 9375    & 50.5\%  & \\\cline{1-5}
     8   & 32   & 10.9\%    & 15000   & 4.4\%   &  \\\hline
    \end{tabular}
    \caption{Experimental results comparing rate (bits/transmission) and probability of semantic error of the two systems at an SNR of 15dB}
    \label{tab_results}
    \vspace{-.3cm}
\end{table}

\vspace{-.2cm}

\printbibliography

\end{document}